\documentclass[12pt]{article}
\makeatletter \oddsidemargin  -.1in \evensidemargin -.1in
\newcommand{\singlespacing}{\let\CS=\@currsize\renewcommand{\baselinestretch}{1}\tiny\CS}
\newcommand{\oneandahalfspacing}{\let\CS=\@currsize\renewcommand{\baselinestretch}{1.25}\tiny\CS}
\newcommand{\doublespacing}{\let\CS=\@currsize\renewcommand{\baselinestretch}{1.35}\tiny\CS}

\newtheorem{rule-def}[theorem]{Rule}

\RequirePackage[dvips]{graphicx} \textheight 22.5cm
\usepackage{amsmath}
\usepackage{subfigure}
\usepackage{graphicx}

\begin{document}

\title{\bf Rheological fluid motion in tube by metachronal waves of
  cilia} 
\author{\small S. Maiti$^{1,2}$ \thanks{Corresponding author, Email address: {\it
      maiti0000000somnath@gmail.com/somnath.maiti@lnmiit.ac.in
      (S. Maiti)}}, S. K. Pandey$^{1}$ \thanks{Email address: {\it
      skpandey.apm@itbhu.ac.in (S. K. Pandey)}}\\ \it $^{1}$Department of Mathematical Sciences\\ \it Indian Institute of Technology (BHU),\\ Varanasi-221005, India\\$^{2}$ Department of Mathematics, The LNM Institute of Information Technology\\Jaipur 302031, India}
\date{}
 \maketitle \noindent \doublespacing
\vspace{-0.5cm}
\begin{abstract}
This paper presents a theoretical study of a non-linear rheological
fluid transport in an axisymmetric tube by cilia. However, an attempt
has been made to explain the role of cilia motion on the transport of
fluid through the ductus efferentes of the male reproductive
tract. Ostwald-de Waele power-law viscous fluid has been considered to
represent the rheological fluid. To analyze pumping by means of a
sequence of beat of cilia from row-to-row of cilia in a given row of
cells and from one row of cells to the next (metachronal wave
movement), we consider the conditions that the corresponding Reynolds
number is small enough for inertial effects to be negligible and the
wavelength to diameter ratio is large enough for the pressure to be
considered uniform over the cross-section.  Analyses and computations
of the detailed fluid motion reveal that the time-averaged flow rate
is dependent on $\epsilon$, a non-dimensional measure involving the
mean radius $a$ of the tube and the cilia length. Thus, flow rate
significantly varies with the cilia length. Moreover, the flow rate
has been reported near to the estimated value $6\times 10^{-3}$$ml/h$
for human efferent ducts if $\epsilon$ is near by 0.4. The estimated
value was suggested by Lardner and Shack \cite{Lardner} in human,
based on the experimental observations for flow rates in efferent
ducts of other animals, e.g., rat, ram and bull. In addition, the
nature of the rheological fluid, i.e., the value of the fluid index
$n$ strongly influences various flow-governed characteristics. The
interesting feature of the paper is that the pumping improves with the
thickening behavior for small values of $\epsilon$ or in free pumping
($\Delta P=0$) and pumping ($\Delta P>0$) regions. \it Chinese Library Classification:{\small O357.2, O361.3
  2010} 
\end{abstract}

\section{Introduction}
The ductus efferentes are a series of microvessels which provide vital
link between the testis and the epididymis \cite{Hess}. It is known
that the wall structure of these vessels is as a single layer of
columnar epithelium which is supported by a thin layer of smooth
muscle and connective tissue \cite{Ilio}. The function of the vessels
are sperm transport from the rete testis to the epididymis and
reabsorbing high amount of fluid which coming from the rete
testis. Due to the second function, spermatozoa concentration raises
to multifold. Epithelium of the ductus efferentes is composed by
ciliated and non-ciliated cells. The well-formed tufts of cilia are
projected by the ciliated cells from the lumen of the duct and
closely-packed, long and regular microvilli are born by the
non-ciliated cells. However, microvilli, which are scattered between
the cilia, are also projected by the ciliated cells; but they are
fewer, shorter and thinner than those of the non-ciliated cells. Since
ductus efferentes are lined by ciliated epithelium, they are unique of
the male reproductive tract \cite{Hess,Ilio}.

A cilium is a hairlike slender appendage/protuberance that projects
from the free surfaces of certain cells (e,g. eukaryotic cells). Its
presence has been observed in almost all animals. Due to their
motility, it engages a significant role in various physiological
processes like locomotion, alimentation, circulation, respiration
and reproduction \cite{Lardner}.

Cilia are classified into motile and non-motile cilia. The latter is
also known as primary cilia. In this study, we have considered motile
cilia which do not beat randomly; but rather in coordinated
manner. This nature of the cilia possesses some important aspects of
ciliated epithelium. A good deal of general observations and
inferences on the cilia of the gill of several aquatic species were
presented by Rivera \cite{Rivera} as: (a) In any given tissue, beat
rate of all the cilia is quite uniform. (b) The lashings of a single
cilium and of cilia on adjacent cells are very much coordinated. (c) A
definite metachronal rhythm is established. Metachronal rhythm is a
movement consisting of a sequence of beat from row to row of cilia in
a given row of cells and then from one row of cells to the next
one. As a result, any object, which is at rest on the surface of cilia,
always moves forwards keeping the direction fixed.

Since metachronal rhythm gives more fixed passage of water with time
over the surface of cilia or perhaps it is impractical to stimulate
synchronous beat of large area (Sleigh \cite{Sleigh1,Sleigh2}), it is believed
that cilia beat in a metachronal rhythm to the contrary of synchronous
manner. However, the metachronal rhythm along the surface of cilia may
alter their pattern (Sleigh \cite{Sleigh1}). The change depends on
whether the metachronal rhythms move towards the effective stroke of
the ciliary beat (symplectic metachronism), or the metachronal rhythms
pass towards the opposite of the effective stroke of the beat and so
in the reverse direction of flow (antiplectic metachronism), or the
cilia beat at right angles to the line of wave movement (diaplectic
metachronism). Satir \cite{Satir2} presented a diaplectic meachronal
rhythm as an example in Figure 6 in his study. There are some data
available for some other animals as wave-lengths, metachronal rhythm
velocities and frequencies (Sleigh \cite{Sleigh1}).

It is a bit debatable as to what extent the cilia and to what extent
smooth muscles drive the fluid in the efferent ducts of the
male reproductive tract. It was reported that the prime contribution
of the net fluid transport in efferent ducts was coming from
metachronal movement of cilia (cf. Lucus \cite{Lucus}, Setchell
\cite{Setchell}). However, Winet \cite{Winet} created a cilio
peristaltic model which suggested that the greater contribution to the
flow was coming from the smooth muscle contraction. Winet \cite{Winet}
remarked that ``we may conclude that if the peristaltic wave has a
33$\%$ or more constriction, the spermatozoa concentration in the
ductus efferentes is at least 4$\times 10^8$ cells $cm^{-3}$ and that
flow rates in the ductus efferentes are the same as in the rete
testis''. However, he added that ``no observations of $\phi$ (the
occlusion factor) have been made for any of the male
tubes''. Epithelium of ductus efferentes is supported by only a thin
layer of smooth muscle and connective tissue (\cite{Ilio}, Page
No. 439, 3rd paragraph). So the suggestion, the major contribution to
the flow was from smooth muscle contraction, is
questionable. Moreover, if the peristaltic wave has a 50$\%$ or more
constriction and cilia length is more than $20\mu m$, cilia of
opposite boundary will clash with each other at the contraction
region. In addition, at the contraction region, the driving force
(which is opposite direction of the flow) at the central region will
be active on the cilia (which obstruct the free space at the central
portion). As a result, there may be damage of cilia indicating that
smooth muscle contraction (i.e. peristaltic motion) may not be the
main driving force of fluid transport at ductus efferentes. Conclusion
of this paper will throw some more light on this issue.

Ilio and Hess \cite{Ilio} stated that the ratio of ciliated and
non-ciliated cells in the epithelium in different animals generally
varies in range between 1:3 and 1:8 (cf. Benoit \cite{Benoit}). So the
vast majority of the inner surface is non-ciliated. However, Aire and
Josling \cite{Aire} reported (based on their experimental
observations) that the cilia of the ciliated cells usually
over-shadowed the luminal surfaces of non-ciliated cells even though
the non-ciliated cells are greater than the ciliated cells in
number (cf. Aire and Josling, pages 194-196, Fig. 9-15). Moreover, in
histological photographs (cf. Aire and Josling \cite{Aire}, Fig. 12,
Page No. 195), it is clear that ciliated and non-ciliated cells are
not distinctly separated into two parts; they are rather adequately
mixed to justify the model considered here.

It is to be noted that data are not easily obtainable for bigger
animal species, especially mammals. However, it was measured that the
beat frequency for cilia lining in the rabbit oviduct was
approximately 20-30 beats/sec (Borell et al. \cite{Borell}). It is
worth mentioning that we can compare this order of magnitude of the
frequency of cilia beat to the lower animals (Sleigh \cite{Sleigh1}).

It is to be further noted that several investigators (c.f. Jahn and
Bovee \cite{Jahn1,Jahn2} and the references therein) studied the
hydrodynamics of protozoa which utilize cilia for locomotion. Blake
\cite{Blake1,Blake2} took a spherical envelope model in order to study
the swimming of the protozoan opalina and the swimming motion
considering the ciliated body either a two-dimensional or cylindrical
shape. Miller \cite{Miller1,Miller2,Miller3} investigated mucus
transport with the help of mechanical simulation of the cilia in the
trachea, while Barton and Raynor \cite{Barton} studied mucus transport
analytically in the trachea without considering the metachronal
rhythm. Those investigations on protozoology and mucus transport in
the respiratory tract gave conception in terms of locomotion in
protozoa as well as movement of particles in the respiratory
tract. However, only a little has been done finding the relation
between the properties of the cilia and the nature of the metachronal
rhythm for fluid movement in efferent ducts. Our motivation came up
from the questions related to the interpretation of the fluid movement
through the ductuli efferentes of male reproductive tract of human
(Greep \cite{Greep}) and the effect of cilia on ovum transport and
sperm movement in fallopian tubes (Blandau \cite{Blandau}), (Sturgis
\cite{Sturgis}). Particularly, this paper will deal with a comparison
of the results for the flow rate of this model to the corresponding
estimated flow rate in the ductuli efferentes of male reproductive
tract.

It is worth to mention that there are limited data available on the
flow rate due to ciliary activity. Based on the experimental
observations (Setchell \cite{Setchell}, Tuck et al. \cite{Tuck},
Waites and Setchell \cite{Waites}), Lardner and Shack \cite{Lardner}
estimated flow rate for human testes as $6\times 10^{-3} ml/hr$ with
approximate values, i.e., $a=50\mu m$ and frequency of beat of the
cilia as 20/sec. But the theoretical model of Lardner and Shack
\cite{Lardner} obtained a flow rate of $0.12\times 10^{-3}
ml/hr$. Hence, further investigations are indeed required.

Past experimental observations indicate that most of the biological
fluids possess non-Newtonian behaviour
\cite{Malek,Dunn,Mendeluk,Xue,Misra1,Misra2,Misra3,Maiti1,Liu,Hayat,Siddiqui1,Ding}. Analyses
on the basis of simple Newtonian fluid yield non realistic
results. The power law model is one of the widely used model for
rheological fluid transport
\cite{Misra1,Misra2,Misra3,Maiti1,Srivastava,Usha,Rao}. The
rheological nature of this model is strongly dependent on the
rheological fluid index $n$. The model approximates both
shear-thinning ($n<1$) and shear-thickening ($n>1$) fluids behaviour
over a large range of rheological conditions
\cite{Malek,Dunn,Misra1,Misra2}. Viscous properties of human semen is
experimentally found to exhibit power law-behaviour (Dunn and
  Picologlu \cite{Dunn}, Mendeluk et al. \cite{Mendeluk}). It has been
reported experimentally that semen proves to fit in a power-law model
with pseudoplastic behavior \cite{Dunn,Mendeluk}. Thus studies on
fluid transport of the power law model by ciliary activity is expected
to yield some important inferences. Constitutive equation of a
Newtonian fluid (i.e., relationship between shear stress and strain
rate) is linear, whereas it is nonlinear for a rheological power-law
fluid.

There are two different approaches for investing the periodic motion
of cilia--the sub-layer model and the envelope model
\cite{Childress}. Sub-layer model approach helps us by computing the
forces and bending moments generated by each cilium to neighbouring
liquid and it obtains the mean flow generated above the cilia layer.
But the hydrodynamic representation is wearisome and normally obtained
numerically \cite{Blake1,Lauga1,Blake3,Vélez-Cordero}. The envelope
model approach has advantage for the consideration of metachronal
rhythms above the cilia layer ignoring the details of the sub-layer
dynamics. Moreover, the envelope model can be utilized for comparing,
even quantitatively, e.g., for comparing the swimming velocities
evaluated mathematically with those reported in water for a number of
microorganisms \cite{Blake2,Vélez-Cordero,Brennen}. In addition, this
approach is accountable to perturbation analysis and has been used to
combine some non-Newtonian effects by a systematic way
\cite{Vélez-Cordero,Lauga2,Lauga3}.

Recently, Siddiqui et al. \cite{Siddiqui2} have studied the flow of a
power law fluid due to ciliary motion in an infinite channel. They
pointed out that the power-law fluid gives results closer to the
estimated one as $6\times 10^{-3} ml/hr$. However, we believe that in
real physiology, the wall contraction/length related to cilia and
favourable pressure gradient less than that they have considered. They
took $\epsilon=0.9$, giving reference to Agarwal and Anawaruddin
\cite{Agarwal} who had reported an application of their model for
fluid transport in vas deferens. It is worth mentioning that ductus
efferentes and vas deferens are quite different ducts, the latter one
may undergo possibly that much wall contraction. $\epsilon$, a
non-dimensional measure with respect to mean radius $a$ of the tube
and the cilia length, would be much less than $\epsilon=0.9$
\cite{Lardner,Blake4}. To the authors' knowledge, the other
unacceptable considerations in \cite{Siddiqui2} are: (i) they
considered a large constant favourable pressure gradient. (ii)
dependence of flow characteristics on higher value of $\delta$, wave
number of the metachronal wave.

It is well known that pressure gradient in small biological vessels
(at least when there is a wave like peristaltic wave
\cite{Sleigh1,Misra2,Satir2,Maiti2,Shapiro}, metachronal wave etc in
the vessels) varies with the length of the vessels and favourable
pressure gradients are physiologically insignificant, in general, and,
at least, in the ductus efferentes (\cite{Winet}). Moreover, study of
flow characteristics is not true for higher values of $\delta$ as they
have done analysis for small values of $\delta$. However, their power
law model is somehow different form and complex than compared with
widely used power-law model.

In view of all the above, a theoretical model is being considered for
more realistic consequences for the flow in an axisymmetric tube under
the influence of metachronal rhythm of cilia movement. We study here a
non-linear problem of Ostwald-de Waele power-law fluid transport
induced by means of a sequence of cilia beat from row-to-row of cilia
in a given row of cells and from one row of cells to the next
(metachronal wave movement). The conditions considered in this study
are small Reynolds number and small wave number.

Based upon the analytical investigation, the velocity, pressure
difference, flow rate have been calculated. Thereafter, we have
carried out extensive numerical calculations. Keeping in view a
specific situation of fluid movement in micro-vessels by ciliary
motion, the numerical results are exhibited in graphs. The results
discussed for shear thinning and shear-thickening fluids are closely
connected in the circumstances of various situations of fluid
transport due to ciliary activity. For example, semen behaviour is
normally of shear-thinning kind when it contains a number of
spermatozoa up to certain limit; however, in the case of heavily
concentrated spermatozoa suspensions, the nature of semen may be
considered as shear-thickening type \cite{Dunn,Mendeluk,Xue}. Moulik
et al. \cite{Moulik} reported that high viscosity of semen might
indicate antibodies in the plasma and/or genital tract infection.

Some novel features are discussed to have a better insight of fluid
motion by ciliary activity. We can utilize ciliary pumping mechanism
in the study of the hydrodynamics of protozoa which use cilia for
locomotion. The results may find useful applications in cilia-based
actuators as micro-mixers for flow control in tiny bio-sensors and
also as micro-pumps for drug-delivery systems.

In this investigation, the term cilia will not consider flagella but
ciliated epithelium. Comparative physiology of flagella including
sperm tails may be available in \cite{Sleigh1,Fawcett}.

\section{Formulation}
A non-linear problem, concerning the fluid movement characteristics in
an axisymmetric tube under the action of ciliary beat that generate a
metachronal rhythm, is to be studied here by considering the fluid as an
incompressible non-Newtonian viscous fluid. We consider an
axisymmetric tube with ciliated walls of large length compared with
its radius and a symplectic metachronal wave is moving in the
righthand side having velocity c. The non-Newtonian viscous fluid
behaviour within the tube is considered to be incompressible
Ostwald-de Waele power law rheological fluid.
\begin{figure}
\centering
\includegraphics[width=4.5in,height=3.0in]{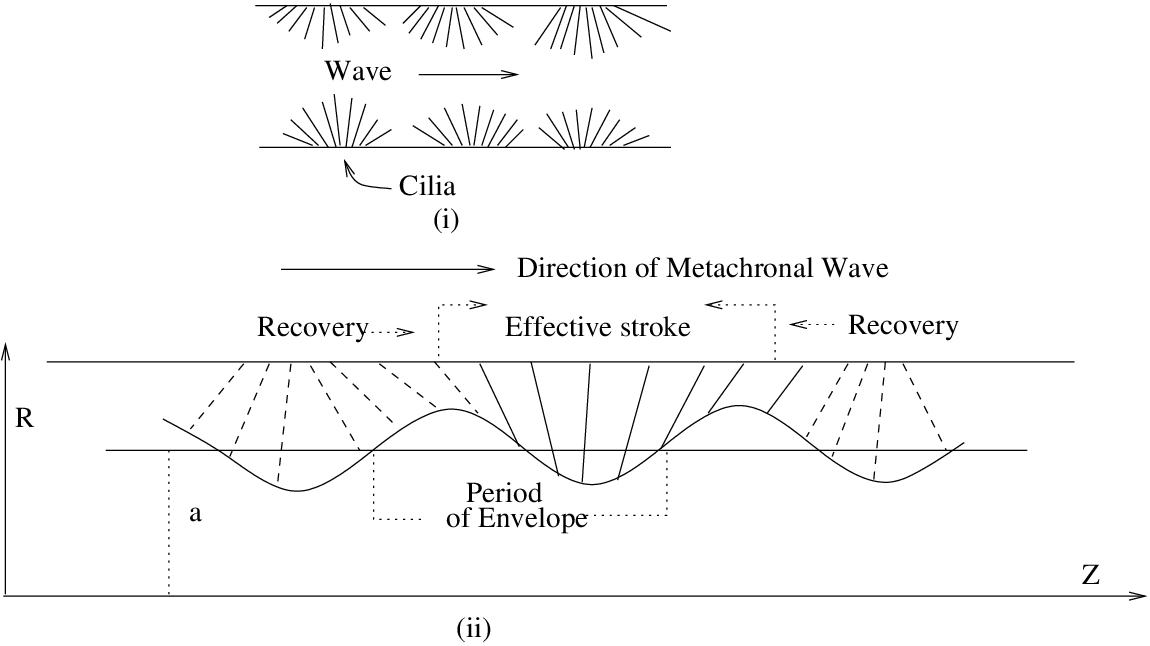}\\\includegraphics[width=4.5in,height=3.0in]{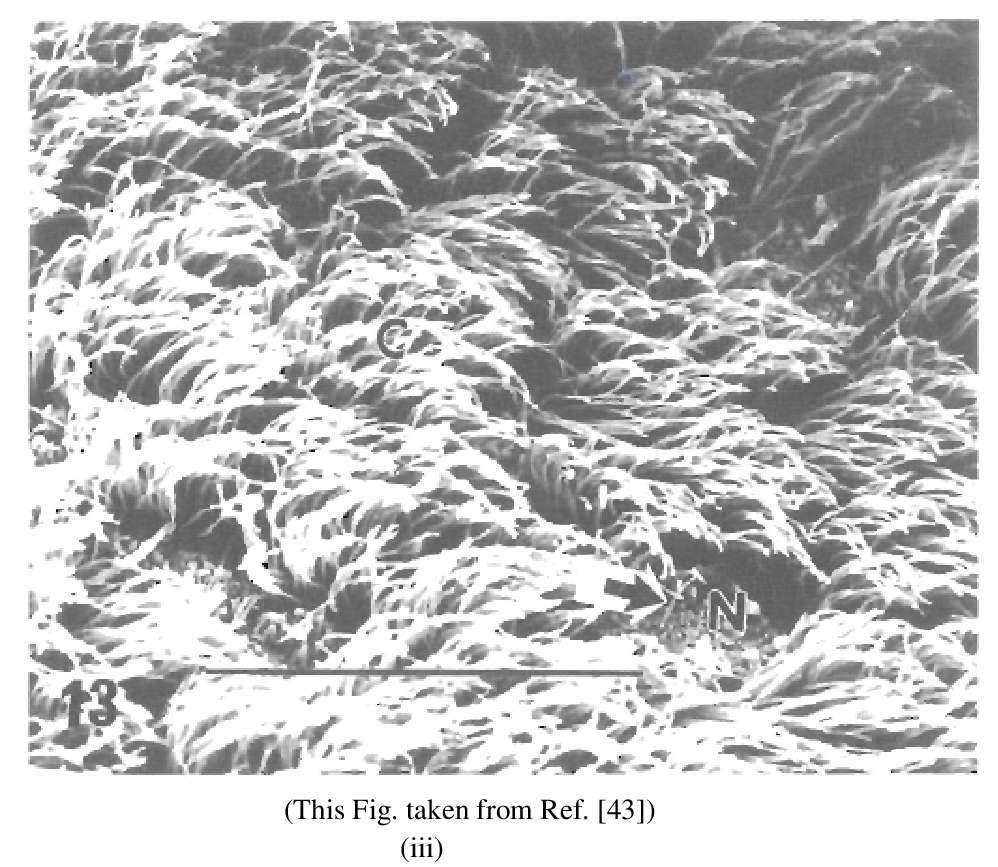}
\caption{Wave motion of Cilia: (i) Ciliated tubule, (ii) metachronal
  wave pattern (iii) a SEM micrograph of the luminal surface of the
  distal efferent duct (DED) of the turkey, showing ciliated (C) , and
  non-ciliated (N) cells. Arrow, single cilium of a non-ciliated
  cell. Bar $=20\mu m$. Perfusion fixation (cf. Ref. \cite{Aire}) }
\label{manuscript_geo7.1}
\end{figure}

We treat (R,$\theta$,Z) as the cylindrical co-ordinates for the position
of any fluid particle. Here, R is measured along the radius of the
tube, the coordinate Z is along the wave propagation direction and
$\theta$ being the rotational coordinate. When $\tau$, $\Delta$ denote the stress
tensor, symmetric rate of deformation tensor respectively, the
governing law of shear stress for the Ostwald-de Waele power
law fluid may be written as \cite{Bird}
\begin{equation}
\tau=-\gamma\left\{\left|\sqrt{\frac{1}{2}(\Delta:\Delta)}\right|^{n-1}\right\}\Delta,
\end{equation}
\begin{eqnarray*}
where~~\frac{1}{2}(\Delta:\Delta)=2\left(\left(\frac{\partial V}{\partial R}\right)^2+\left(\frac{V}{R}\right)^2+\left(\frac{\partial U}{\partial Z}\right)^2\right)+\left(\frac{\partial U}{\partial R}+\frac{\partial V}{\partial R}\right)^2,
\end{eqnarray*}
where $\gamma$, $n$ are respectively the flow consistency index and
power-law index respectively. We introduce $U$ and $V$ as the velocity
components respectively in the $Z$, $R$ directions. We know that shear
thinning fluid has been classified for $n<1$ and $n>1$ corresponds to
shear thickening fluid.

Under the above consideration, the flow of an incompressible
viscous Ostwald-de Waele power-law fluid in an axisymmetric tube
together with equation of continuity may be considered to be governed
by
\begin{equation}
\frac{1}{R}\frac{\partial
  (RV)}{\partial R}+\frac{\partial
U}{\partial Z}=0,
\end{equation}
\begin{equation}
\rho \left (\frac{\partial U}{\partial t}+U\frac{\partial
U}{\partial Z}+V\frac{\partial U}{\partial R}\right
)=-\frac{\partial P}{\partial Z}-\frac{1}{R}\frac{\partial
  (R\tau_{RZ})}{\partial R}-\frac{\partial \tau_{ZZ}}{\partial Z},
\end{equation}
\begin{equation}
\rho\left (\frac{\partial V}{\partial t}+U\frac{\partial V}{\partial
  Z}+V\frac{\partial V}{\partial R}\right )=-\frac{\partial
  P}{\partial R}-\frac{1}{R}\frac{\partial (R\tau_{RR})}{\partial
  R}-\frac{\partial \tau_{RZ}}{\partial Z},
\end{equation}
where $\rho$ and $P$ are the density and pressure of the fluid respectively.
Geometry of the metachronal wave pattern helps us to assume the envelope of the cilia tips mathematically by
\begin{eqnarray}
R=H=f(Z,t)=a+a\epsilon~\cos\left(\frac{2\pi}{\lambda}(Z-ct)\right).
\label{manuscript_envelope_cilia_tip}
\end{eqnarray}
This can be therefore assumed as the equation of the extensible vessel
wall. Here `$a$' is mean radius of the tube and $\epsilon$ be a
non-dimensional parameter which together with `$a$' in the form of
`$a\epsilon$' makes the amplitude of the metachronal wave.  $\lambda$,
$c$ are the metachronal wave length and velocity
respectively. Observations of the various patterns of cilia movement
in \cite{Sleigh2} motivate us to assume the cilia tips to move in
elliptical paths so that the horizontal position of a cilia tip is in
an implicit form
\begin{eqnarray}
Z=g(Z,Z_0,t)=Z_0+a\epsilon\alpha~\sin\left(\frac{2\pi}{\lambda}(Z-ct)\right),
\label{manuscript_envelope_cilia_tip_horizontal}
\end{eqnarray}
in which $Z_0$ represents a reference location of the particle,
$\alpha$ stands for a measure of the eccentricity of the elliptical
motion. Under the no slip condition, the velocities imparted to fluid
particles are just those of the cilia tips. Thus, the axial and
vertical velocities of the cilia are evaluated as
\begin{eqnarray}
U=\frac{\partial Z}{\partial t}|_{Z_0}=\frac{\partial g}{\partial t}+\frac{\partial g}{\partial Z}\frac{\partial Z}{\partial t}=\frac{\partial g}{\partial t}+\frac{\partial g}{\partial Z}U,
\label{manuscript_cilia_velo_horizontal_equ}\\
V=\frac{\partial R}{\partial t}|_{Z_0}=\frac{\partial f}{\partial t}+\frac{\partial f}{\partial Z}\frac{\partial Z}{\partial t}=\frac{\partial f}{\partial t}+\frac{\partial f}{\partial Z}U.
\label{manuscript_cilia_velo_vertical_equ}
\end{eqnarray}
If the equations (\ref{manuscript_envelope_cilia_tip}) and
(\ref{manuscript_envelope_cilia_tip_horizontal}) are applied to
the equations (\ref{manuscript_cilia_velo_horizontal_equ}) and
(\ref{manuscript_cilia_velo_vertical_equ}), we obtain
\begin{eqnarray}
U=\frac{\frac{-2\pi ac\alpha\epsilon}{\lambda}\cos\left(\frac{2\pi}{\lambda}(Z-ct)\right)}{1-\frac{2\pi a\alpha\epsilon}{\lambda}\cos\left(\frac{2\pi}{\lambda}(Z-ct)\right)}
\label{manuscript_cilia_velo_horizontal}\\
V=\frac{\frac{2\pi ac\epsilon}{\lambda}\sin\left(\frac{2\pi}{\lambda}(Z-ct)\right)}{1-\frac{2\pi a\alpha\epsilon}{\lambda}\cos\left(\frac{2\pi}{\lambda}(Z-ct)\right)}
\label{manuscript_cilia_velo_vertical}
\end{eqnarray}
These boundary conditions enable us to distinguish between the
effective stroke of the cilia and the slow less effective recovery
stroke by approximately accounting for the shortening of the
cilia. Hence, the tube is narrower when $U$ is positive at the
boundary.

For a wave frame ($z$,$r$) moving with a velocity c away from
a fixed frame ($Z$,$R$), let us apply the transformations
\begin{equation}
z=Z-ct, ~~r=R, ~~ u=U-c, ~~ v=V, ~~ p(z,t)=P(Z,R,t),
\label{manuscript_transformation}
\end{equation}
where ($u$,$v$), ($U$,$V$) denote the velocity components, $p$ and $P$
are respectively pressure in wave frame and fixed frame of
reference. Afterward, we will introduce the following
non-dimensional variables:
\begin{eqnarray}
\bar{z}=\frac{z}{\lambda},~~\bar{r}=\frac{r}{a},~~\bar{u}=\frac{u}{c},~\bar{v}=\frac{v}{c\delta},~
\delta=\frac{a}{\lambda},~\bar{p}=\frac{a^{n+1}p}{\gamma
  c^n\lambda},~\bar{t}=\frac{ct}{\lambda},
~h=\frac{H}{a},\nonumber\\ Re=\frac{\rho
  a^n}{\gamma
  c^{n-2}}\frac{a}{\lambda},~\bar{\tau}_{rz}=\frac{\tau_{rz}}{\gamma\left(\frac{c}{a}\right)^n},~\bar{Q_1}=\frac{Q_1}{\pi a^2c},~~~~~~~~
\label{manuscript_non-dimensionalize}
\end{eqnarray}
where $Q_1(Z,t)$ is the instantaneous volume flow rate and $Re$ is
Reynolds number of the fluid flow. Dropping the bars over the symbols,
the equations governing the flow can be rewritten as
\begin{equation}
\frac{1}{r}\frac{\partial (rv)}{\partial
  r}+\frac{\partial u}{\partial z}=0,
\label{manuscript_continuity_non-dimensional}
\end{equation}
\begin{equation}
 Re\left (\frac{\partial u}{\partial t}+u\frac{\partial
u}{\partial z}+v\frac{\partial u}{\partial r}\right
)=-\frac{\partial p}{\partial z}+\frac{1}{r}\frac{\partial
  \left(\Phi\left(r\frac{\partial u}{\partial r}+r\delta^2\frac{\partial v}{\partial
z}\right)\right)}{\partial r}+2\delta^2\frac{\partial \left(\Phi\frac{\partial
u}{\partial z}\right)}{\partial z},
\end{equation}
\begin{equation}
Re\delta^2\left (\frac{\partial v}{\partial t}+v\frac{\partial v}{\partial
z}+v\frac{\partial v}{\partial r}\right )=-\frac{\partial
p}{\partial r}+\delta^2\frac{1}{r}\frac{\partial
  (r\Phi\frac{\partial v}{\partial r})}{\partial r}+\delta^2\frac{\partial\left(\Phi(\frac{\partial u}{\partial r}+\delta^2\frac{\partial v}{\partial
z})\right)}{\partial z},
\end{equation}
\begin{eqnarray}
\Phi=\left|\sqrt{2\delta^2\left\{\left(\frac{\partial v}{\partial r}\right)^2+\left(\frac{v}{r}\right)^2+\left(\frac{\partial
u}{\partial z}\right)^2\right\}+\left(\frac{\partial u}{\partial r}+\delta^2\frac{\partial v}{\partial
z}\right)^2}\right|^{n-1};
\end{eqnarray}
whereas the boundary conditions are 
\begin{eqnarray}
\frac{\partial u}{\partial r}=0 \rm{~at~}r=0 \rm{~and~} u=\frac{-2\pi \alpha\delta\epsilon\cos (2\pi z)}{1-2\pi \alpha\delta\epsilon\cos (2\pi z)}-1,
\label{manuscript_cilia_velo_horizontal_dimensionless}\\
v=\frac{2\pi\epsilon \sin (2\pi z)}{1-2\pi \alpha\delta\epsilon\cos (2\pi z)},
\label{manuscript_cilia_velo_vertical_dimensionless}
\end{eqnarray}
on $r=h=1+\epsilon\cos (2\pi z)$.

Since in most of the cases of flow in small diameter tubules, Reynolds
numbers are very small ($Re\ll 1$), the analysis can be carried out by
the approximation of the inertia-free flow. Moreover, in the wave
frame of reference, if the tube length is finite and equal to an
integral number of wavelengths together with constant pressure
difference across the ends of the tube, the flow may be steady
\cite{Shapiro}. In this case, the governing equations under the
consideration of the long-wavelength approximation (which makes
$\delta <<1$) can be simplified as follows:
\begin{equation}
0=-\frac{\partial p}{\partial z}+ \frac{1}{r}\frac{\partial
  (r\frac{\partial u}{\partial r}|\frac{\partial u}{\partial
    r}|^{n-1})}{\partial r}
\label{manuscript_zmomentum_lubrication}
\end{equation}
\begin{equation}
0=-\frac{\partial p}{\partial r}
\label{manuscript_rmomentum_lubrication}
\end{equation}
and also the simplified forms of the boundary conditions may be given
by
\begin{eqnarray}
\frac{\partial u}{\partial r}=0 \rm{~at~}r=0 \rm{~and~} u=u(h)=-1-2\pi \alpha\delta\epsilon\cos (2\pi z),
\label{manuscript_cilia_velo_horizontal_simplified}\\
v=v(h)=2\pi \epsilon\sin (2\pi z)+2\pi^2\epsilon^2\alpha\delta\sin (4\pi z),
\label{manuscript_cilia_velo_vertical_simplified}
\end{eqnarray}
on $r=h=1+\epsilon\cos (2\pi z)$. By solving
(\ref{manuscript_zmomentum_lubrication}) subject to the boundary
conditions (\ref{manuscript_cilia_velo_horizontal_simplified}), we find the axial velocity in the form
\begin{equation}
u(r,z)=u(h)+\frac{p_1|p_1|^{k-1}}{2^k(k+1)}\left[r^{k+1}-h^{k+1}\right],
\label{manuscript_cilia_axial_velocity}
\end{equation}
where $p_1=\frac{\partial p}{\partial z}$ and $k=\frac{1}{n}$. If we integrate the continuity (\ref{manuscript_continuity_non-dimensional}) across the cross section of the tube, we get
\begin{eqnarray}
&&2\int_{0}^{h}\frac{1}{r}\frac{\partial (rv)}{\partial r}rdr+2\int_{0}^{h}\frac{\partial u}{\partial
    z}rdr=0\\&&\textrm{or,~} rv|_h-rv|_0+\frac{\partial}{\partial
    z}\int_{0}^{h}urdr-hu(h,z,t)\frac{\partial h}{\partial z}-0\times
  u(0,z,t)=0\\&&{\rm or,~}
hv(h)+\frac{1}{2}\frac{\partial q}{\partial z}-h\frac{\partial h}{\partial z}u(h)=0, \textrm{where~}q=2\int_{0}^{h}urdr
\end{eqnarray}

\begin{eqnarray}
\textrm{which~ implies~ that~}\frac{\partial q}{\partial z}=2h\left(\frac{\partial h}{\partial z}u(h)-v(h)\right)=0.~~~~~~~~~~~~~~~~~~~~~~~~~~~~~~~~~~~~~~~~~~~~~~~~~~~~~~~~~~~~~~~~~~~~~~~~~~~~~~~~~~~~~~~~~~~~~~~~~~~~~~~~~~~~~~~~~~~~~~~~~~~~~~~~~~~~~~~~~~~~~~~~~~~~~~~~~~~~~~~~~~~~~~~~
\label{manuscript_cilia_flow_rate}
\end{eqnarray}
This shows the fixed volume flow rate $q$ in the wave frame of
reference. By integrating (\ref{manuscript_cilia_axial_velocity})
across the cross section of the tube, pressure gradient is obtained in
terms of the volume flow rate in the following form:
\begin{equation}
q=h^2u(h)-\frac{p_1|p_1|^{k-1}h^{k+3}}{2^k(k+3)}.
\label{manuscript_cilia_flow_rate_pressure_grad}
\end{equation}
Putting $dp/dz$, which obtained from
(\ref{manuscript_cilia_flow_rate_pressure_grad}), into
(\ref{manuscript_cilia_axial_velocity}) we calculate
\begin{equation}
u(r,z)=u(h)+\frac{(k+3)(q-h^2u(h))}{(k+1)h^{k+3}}\left[h^{k+1}-r^{k+1}\right]
\label{manuscript_cilia_axial_velocity_containing_flow_rate}
\end{equation}
By integrating, the instantaneous non-dimensional volume flow rate,
$Q_1(Z,t)$, in the fixed frame of reference, is found as
\begin{equation}
Q_1(Z,t)=2\int_{0}^{h}U(R,Z,t)RdR=q+h^2(\rm{~using ~the~
  transformation~formulae~ (\ref{manuscript_transformation})})
\label{manuscript_cilia_flow_rate_fixed_frame}
\end{equation}
Thus, the time-mean volume flow rate over a wave period is given by
\begin{equation}
Q(Z)=\frac{1}{T}\int_{0}^{T}Q_1(Z,t)dt=q+\frac{1}{T}\int_{0}^{T}h^2(Z,t)dt,
\label{manuscript_cilia_flow_rate_time_averaged_fixed_frame}
\end{equation}
which, on integration for the sinusoidal wall of (\ref{manuscript_envelope_cilia_tip}), gives
\begin{equation}
Q(Z)=q+1+\frac{\epsilon^2}{2}
\label{manuscript_cilia_time_averaged_flow_rate_fixed_frame}
\end{equation}
Solving $dp/dz$ from
(\ref{manuscript_cilia_flow_rate_pressure_grad}) and using
(\ref{manuscript_cilia_time_averaged_flow_rate_fixed_frame}),
the pressure rise per wavelength may be given by the relation
\begin{equation}
\Delta
p=\int_{0}^{1}\frac{dp}{dz}dz=-\int_{0}^{1}\left|\frac{2^k(k+3)(Q-1-\frac{\epsilon^2}{2}-h^2u(h)}{h^{k+3}}\right|^{n-1}\left\{\frac{2^k(k+3)(Q-1-\frac{\epsilon^2}{2}-h^2u(h))}{h^{k+3}}\right\}dz.
\label{manuscript_cilia_pressure_rise_relation}
\end{equation}
It may be noted that if we set $n=1$, $\alpha=0$ in Equations
(\ref{manuscript_cilia_axial_velocity}),
(\ref{manuscript_cilia_flow_rate_pressure_grad}) and
(\ref{manuscript_cilia_pressure_rise_relation}), the expressions
reduce to those reported earlier in Ref. \cite{Shapiro}. If
$\alpha=0$, the results also match well those of Ref. \cite{Misra1}
when the peristaltic wave form in \cite{Misra1} is replaced by the
metachronal wave in this study. Similarly, if the yield stress is set
equal to zero and fluid flow is considered in an axisymmetric uniform
vessel in \cite{Maiti2}, the expressions said above reduce to that
obtained in \cite{Maiti2}.

By solving the continuity equation
(\ref{manuscript_continuity_non-dimensional}) subject to the boundary
condition (\ref{manuscript_cilia_velo_vertical_simplified}), we get
the radial velocity as
\begin{equation}
v(r,z)=u(h)\frac{r^{k+2}}{h^{k+2}}\frac{\partial h}{\partial
   z}+\frac{h}{k+1}\left[\frac{r}{h}-\frac{r^{k+2}}{h^{k+2}}\right]\frac{\partial u(h)}{\partial
   z}+\frac{(k+3)q}{(k+1)h^2}\left[\frac{r}{h}-\frac{r^{k+2}}{h^{k+2}}\right]\frac{\partial h}{\partial
   z}.
\label{manuscript_cilia_radial_velocity}
\end{equation}

\begin{figure}
\centering
\includegraphics[width=3.35in,height=1.8in]{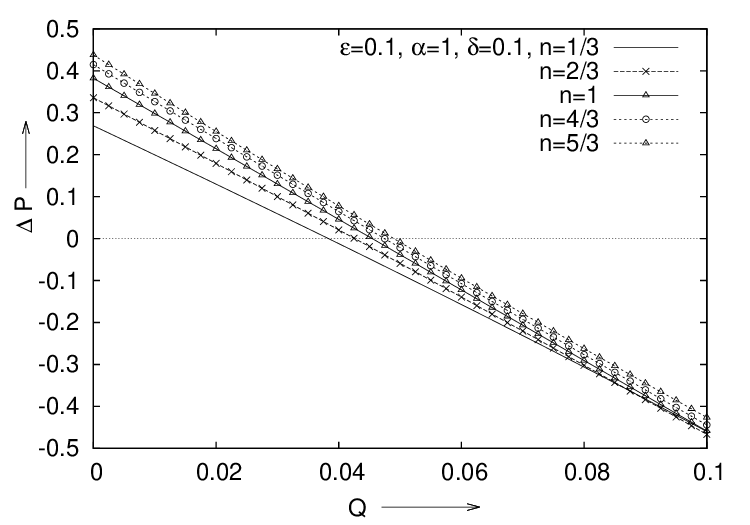}\includegraphics[width=3.35in,height=1.8in]{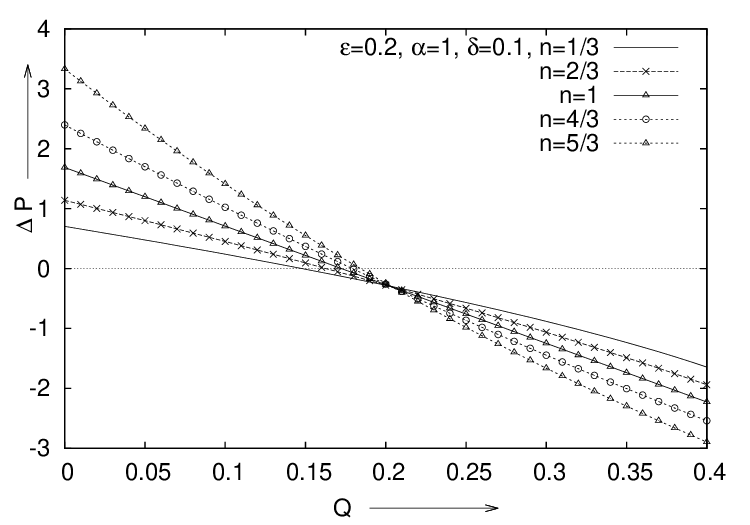}\\$~~~~~~~~~~~~~~~~~~~~~~~~~~~(a)~~~~~~~~~~~~~~~~~~~~~~~~~~~~~~~~~~~~~~~~~~~~~~~~~~~~~~~~~~~~~~~~~~~~~(b)~~~~~~~~~~~~~~~$
\includegraphics[width=3.35in,height=1.8in]{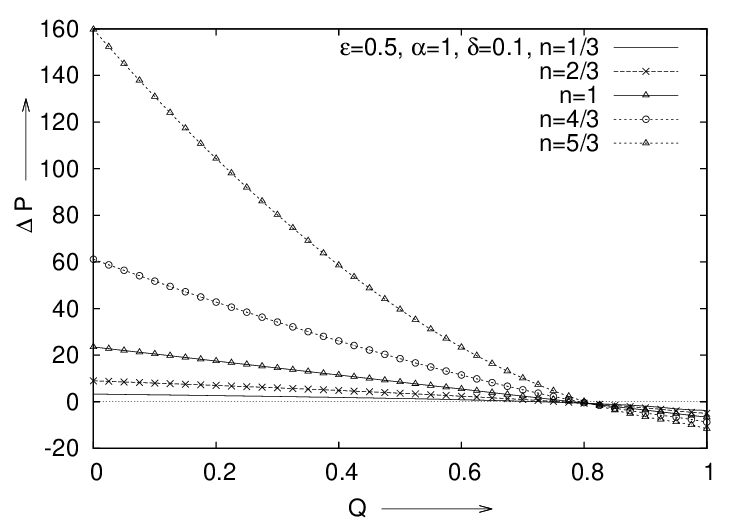}\\(c)
\caption{The diagrams exhibit the dependence of flow rate and pressure difference relation ($Q-\Delta p$) on $\epsilon$, flow behaviour index $n$ and eccentricity $\alpha$. In all the figures, wave number $\delta=0.1$, $\alpha=1$ and $n$ has been varied from 1/3 to 5/3 to observe the changes taking place in $Q-\Delta p$ relation. Effect of $\epsilon$ on $Q-\Delta p$ relation has been shown in Figs. (a-c) by varying $\epsilon$ from 0.1-0.5.}
\label{manuscript_pump7.1.1-7.4.1}
\end{figure}

\begin{figure}
\centering
\includegraphics[width=3.35in,height=1.8in]{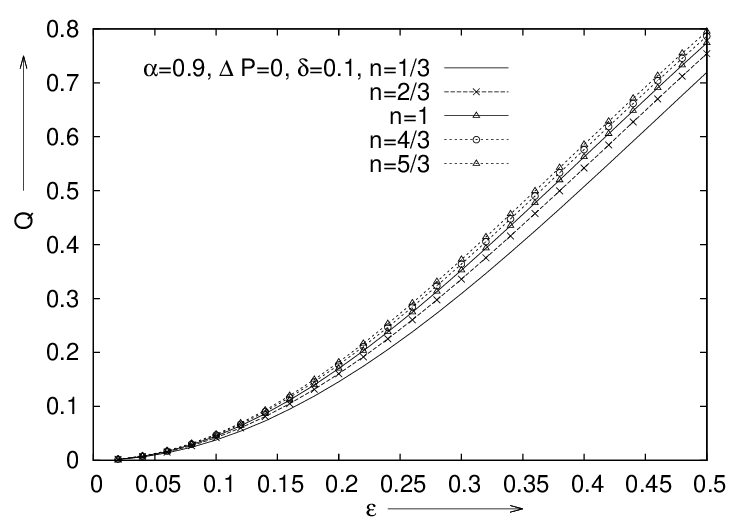}
\caption{The diagram exhibits the dependence of flow rate $Q$ and
  $\epsilon$ relation on flow behaviour index $n$ by plotting graphs
  for n=1/3-5/3. The other parameters are kept as $\Delta p=0$,
  i.e. free pumping, $\delta=0.1$, $\alpha=0.9$.}
\label{manuscript_pump7.2.1}
\end{figure}

\begin{figure}
\centering
\includegraphics[width=3.35in,height=1.8in]{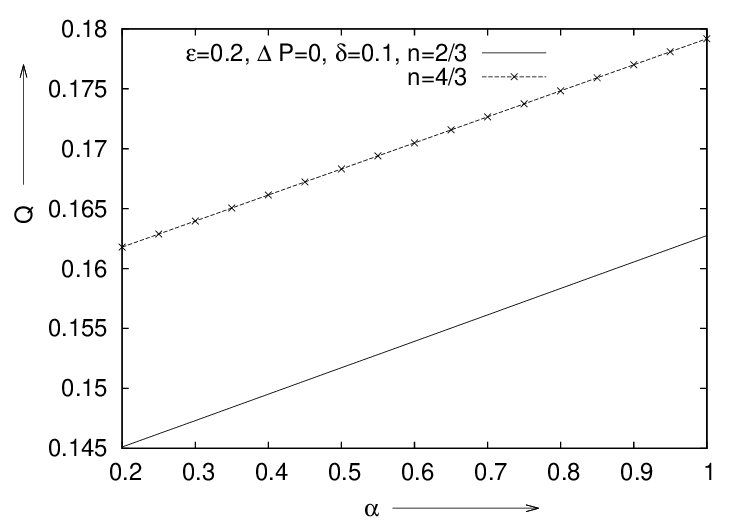}
\caption{The diagram exhibits the dependence of flow rate $Q$ and
  eccentricity $\alpha$ relation on flow behaviour index $n$ by
  plotting graphs for n=2/3-4/3. The other parameters are kept as
  $\Delta p=0$, i.e. free pumping, $\delta=0.1$, $\epsilon=0.2$.}
\label{manuscript_pump7.3.1}
\end{figure}

\begin{figure}
\centering
\includegraphics[width=3.35in,height=1.8in]{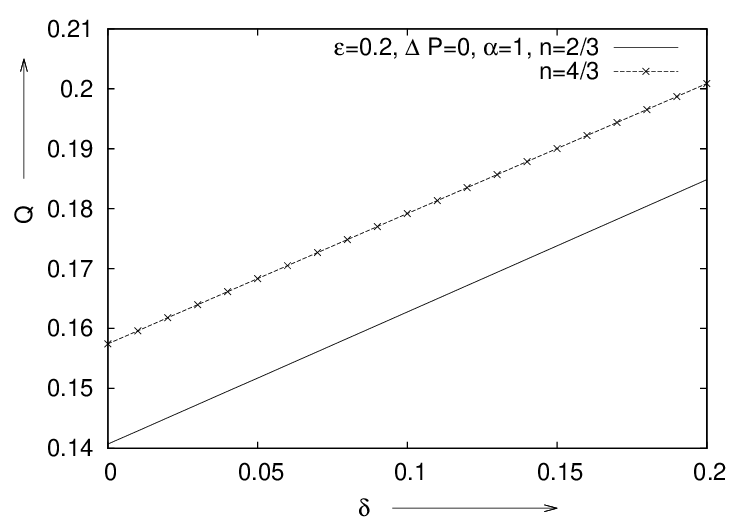}
\caption{The diagram exhibits the relation of flow rate $Q$ with wave
  number $\delta$ and its dependence on flow behaviour index $n$,
  which has been shown by plotting graphs for n=2/3 and 4/3. The other
  parameters are kept as $\Delta p=0$, i.e. free pumping, $\alpha=1$,
  $\epsilon=0.2$.}
\label{manuscript_pump7.4.1}
\end{figure}

\section{Quantitative Study}
This section deals with a quantitative study of the mathematical model
considered in the above sections. It may be noted that unlike the
Newtonian model, it has been not possible to find an explicit
expression for pressure difference $\Delta p$ in terms of time
averaged flow rate $Q$ for Ostwald-de Waele power law model. This is
due to the complexity of the problem of a rheological fluid arising
out of arbitrary wave shape. However, a numerical computation is
required to express pumping characteristics or pumping performance of
the rheological fluid for the sake of clarity. In addition, it is a
convention to exhibit velocity field in term of pressure difference
between the tube ends. Hence, in view of those observations and also
in order to fulfill the requirements, necessary computations have been
carried out numerically by using the software MATHEMATICA.

It is to be noted that there are only a few available data on the flow
rates due to ciliary activity \cite{Lardner,Blake4}. For the
quantitative study, we shall present mathematical estimates of various
physical quantities relevant to physical problems of the flows of
rheological fluids under ciliary activity. For the present analysis,
the following non-dimensional data for the rheological fluids have
been used \cite{Lardner,Blake4}: \\ $\epsilon=0.1$~ to 0.5, $\alpha=$
0.3 to 1, $\delta=0.05$ to 0.2, $n=\frac{1}{3}$ to 2, $Q=0$ to 1.

\subsection{Pumping Characteristics}
It is natural to determine pressure-flow characteristic (i.e. the
pumping characteristics) by the variation of time averaged flow rate
$Q$ with pressure difference$\Delta P$ across one wave
length. Variations of volumetric flow rate by cilia motion for
different values of $\epsilon$, flow index number $n$, wave number
$\delta$ and eccentricity $\alpha$ have been exhibited in
Figs. \ref{manuscript_pump7.1.1-7.4.1}-\ref{manuscript_pump7.4.1}. As
expected, the variations are linear nature for an inertia-free flow
($Re$=0) of a Newtonian fluid. The curves of $\Delta P$ versus $Q$ are
straight lines with negative slope and positive intercepts; but,
for the aforesaid case of a rheological fluid, the relation is
nonlinear between the pressure difference and the mean flow rate.

The curves have three portions, namely, (i) free pumping zone,
describing the region in which $\Delta P=0$, (ii) pumping zone,
indicating the region where $\Delta P>0$ and (iii) co-pumping zone,
which is regarded for the region where $\Delta P<0$, the situation
favourable for the flow to take place. We can compute the amount of
flow pumped by ciliary activity if the mean pressure gradient ($\Delta
P=0$) is absent and also for the case of adverse pressure gradient
(i.e $\Delta P>0$) up to a certain limit. If $\epsilon$, $n$,
$\delta$, $\alpha$ are fixed at some values, employment of an adverse
mean pressure gradient diminishes the mean flow rate obtained when
$\Delta P=0$; and by the time it becomes right amount (certain limit),
it is equal to the driving force of motion generated by the cilia
movement, i.e. the mean flow reduces to 0. Further, if $\Delta P$
($>0$) exceeds the limit, the fluid will move in the reverse direction.

Figs. \ref{manuscript_pump7.1.1-7.4.1}(a-c) show that the area of the
pumping zone and the length of the free pumping zone both increase
significantly with a rise in $\epsilon$ for Newtonian, shear thinning
as well as shear thickening fluids. Also these (area, length) enhance
at a very high rate with an increase in the rheological parameter
$n$. In other words, for fixed values of $\Delta P$, time averaged
flow rate $\bar{Q}$ rises with an increase in $\epsilon$ and
$n$. However, in the co-pumping zone, there is a critical value of
$\Delta P$ for which $Q$ can be raised for all fixed values of
$\Delta P$ with an increase in $n$ when $\epsilon$ $>0.1$ . If $\Delta P$
exceeds this critical value, the reverse trend occurs. It is important
to mention that for fixed values of parameters, taken by Lardner and
Shack together with $n=1$, the flow rate in axisymmetric tube is twice
the values reported by Lardner and Shack for a two-dimensional
channel. In the case of free pumping, 
Fig. \ref{manuscript_pump7.2.1} clearly reveals that there is
a remarkable enhancement of flow rate with increase in $\epsilon$. It
further explains that $n$ strongly influences the flow rate. The
plots, presented in Figs. (\ref{manuscript_pump7.3.1}-\ref{manuscript_pump7.4.1}),
indicate that for a non-Newtonian fluid, $Q$ increase as the wave
number $\delta$ and $\epsilon$ increase and these parameters enhance
$Q$ at almost the same rate for both the shear thinning and shear
thickening fluids.

It is worthwhile to mention that an increase of $\epsilon$ corresponds
to rise of cilia length and vice versa. Again in the case of free
pumping, when $\epsilon$=0, i.e., there are no cilia to the inner
surface of the tube, then $Q$=0.

\begin{figure}
\centering
\begin{minipage}[b]{0.47\textwidth}
\centering
\subfigure[Axial velocity at t=0]{
\label{paper3_fig3.1}
\includegraphics[width=3.6in]{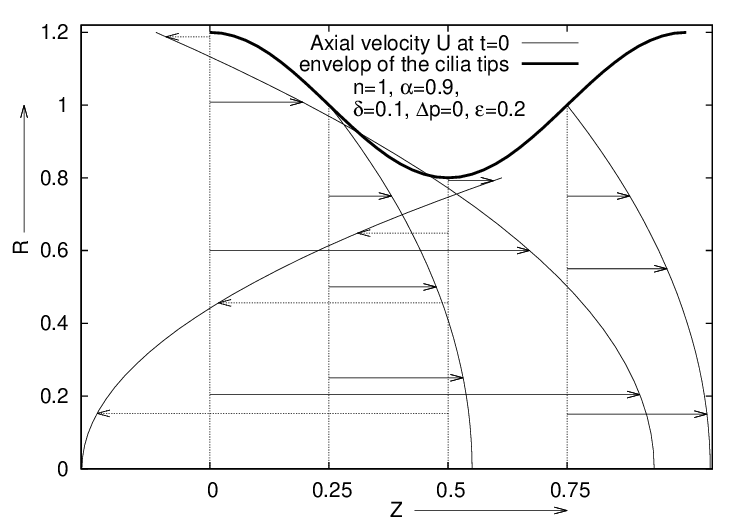}}
\end{minipage}\hspace{0.3in}
\begin{minipage}[b]{0.47\textwidth}
\centering
\subfigure[Axial velocity at t=0.25]{
\label{paper3_fig3.2}
\includegraphics[width=3.6in]{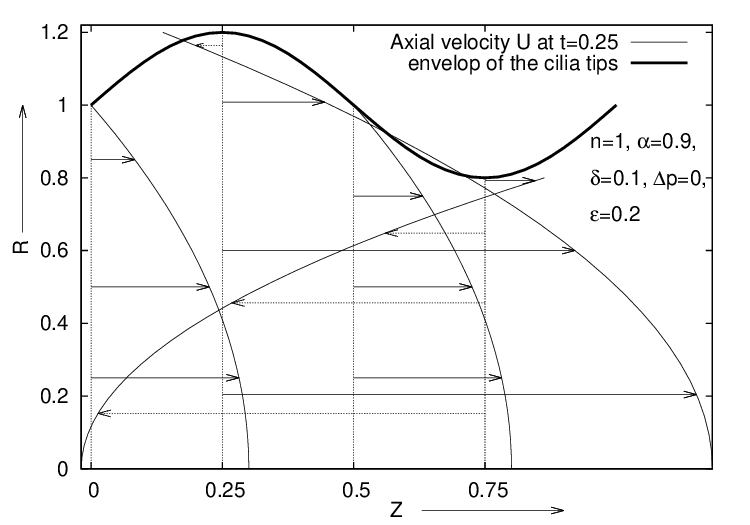}}
\end{minipage}
\\
\begin{minipage}[b]{0.47\textwidth}
\centering
\subfigure[Axial velocity at t=0.5]{
\label{paper3_fig3.3}
\includegraphics[width=3.6in]{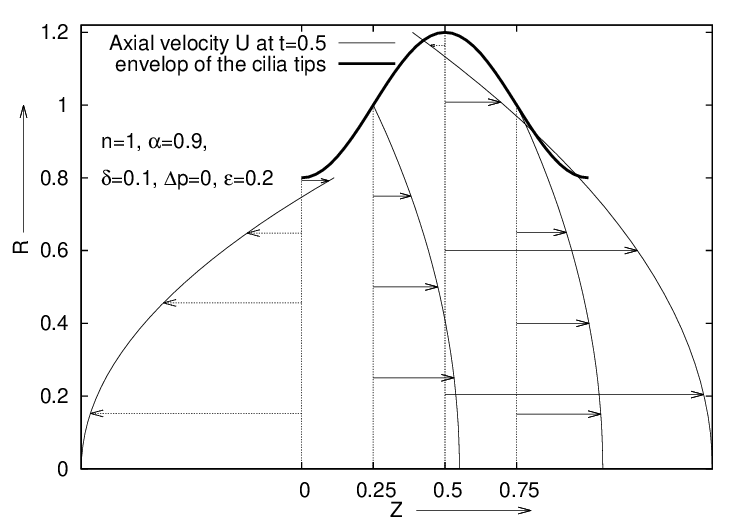}}
\end{minipage}\hspace{0.3in}
\begin{minipage}[b]{0.47\textwidth}
\centering
\subfigure[Axial velocity at t=0.75]{
\label{paper3_fig3.4}
\includegraphics[width=3.6in]{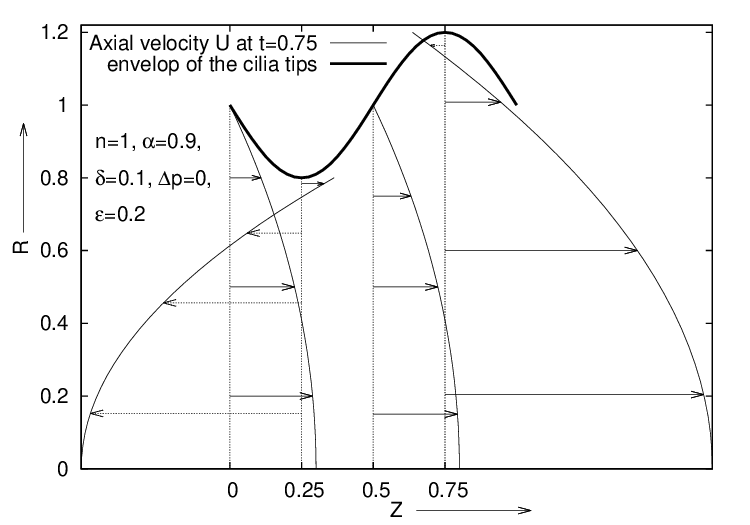}}
\end{minipage}
\caption{The dark coloured curves represent envelop of cilia tips in
  the diagrams. Within them and the central lines of the tubes are
  given axial positions, represented by vertical lines, of fluid
  particles considered at different instants of time ($t=0-0.75$)
  depicted in 4 different diagrams. The arrows emerging from the
  vertical lines indicate their velocities and directions which
  eventually form profiles indicated by curves touching the heads of
  the arrows. The arrows in the backward direction indicate retrograde
  flows.}
\label{manuscript_velo7.1.1-7.1.4}
\end{figure}

\subsection{Distribution of Velocity}
As shown in the previous section, the mean flow rate is only
responsible to the flow through ciliary activity during free pumping
and only the movement of cilia produces the driving force for fluid
transport. For various values of $\epsilon$, $n$, $\delta$ and
$\alpha$, the distributions of axial velocity of the current
investigation are presented in
Figs. (\ref{manuscript_velo7.1.1-7.1.4},
\ref{manuscript_velo7.2.1-7.2.6}-\ref{manuscript_velo7.5.1-7.5.13}). Since
the flow is unsteady in the fixed frame of reference and the velocity
profiles along with the lumen of the ductus efferentes vary with time,
the distribution of velocity has been investigated at a time interval
which is a quarter of a metachronal wave period of cilia.

\begin{figure}
\centering
\includegraphics[width=6.0in,height=4.0in]{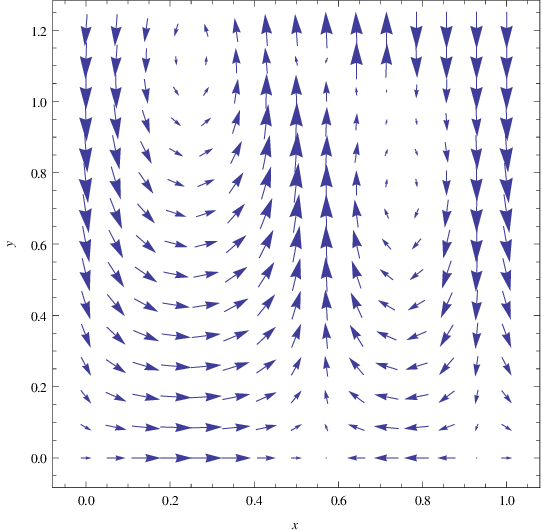}
\caption{The diagram shows instantaneous flow field at time t=0.25 of a
  Newtonian fluid ($n=1$) when $\epsilon=0.2$, $\Delta P=0$, $\alpha=0.9$, $\delta=0.1$.}
\label{manuscript_velovect}
\end{figure}

\begin{figure}
\centering
\begin{minipage}[b]{0.47\textwidth}
\centering
\subfigure[Velocity distribution of Newtonian fluid at \newline the wave crest]{
\label{fig7.2.1}
\includegraphics[width=3.6in]{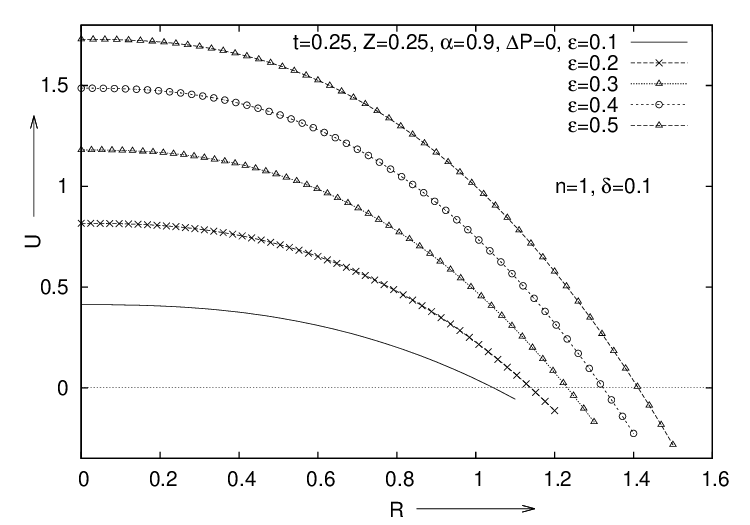}}
\end{minipage}\hspace{0.3in}
\begin{minipage}[b]{0.47\textwidth}
\centering
\subfigure[Velocity distribution of Newtonian fluid at the wave trough]{
\label{fig7.2.6}
\includegraphics[width=3.6in]{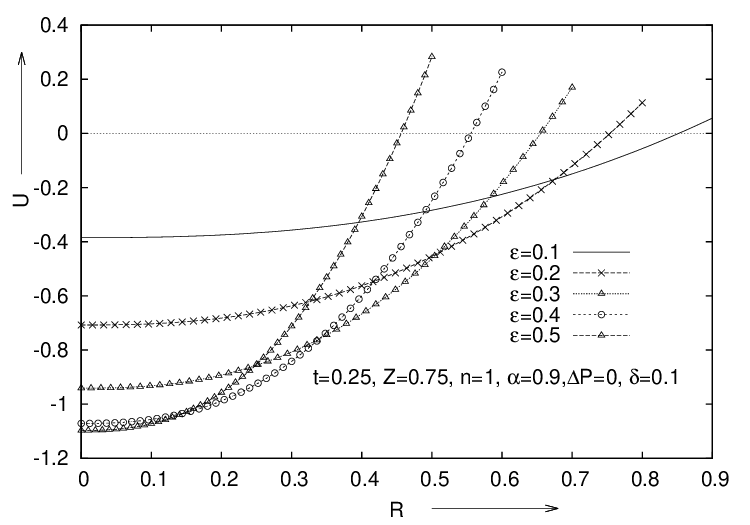}}
\end{minipage}
\caption{The diagrams show the distributions of velocities of a
  Newtonian fluid at the wave crest and the wave trough section for
  the different values of $\epsilon$ when $\Delta P=0$.}
\label{manuscript_velo7.2.1-7.2.6}
\end{figure}

One can observe from Fig.  \ref{manuscript_velo7.1.1-7.1.4} that there
exists a retrograde flow region at any time and the maximum retrograde
flow occurs at the narrowest portion of the tube, while the maximum
flow in the forward direction takes place at the the widest portion of
the tube. As the time-averaged flow rate is positive for free-pumping
(cf. Pumping Characteristic in the previous section), the forward flow
region is predominant here. Moreover, the study further reveals the
existence of two stagnation points on the axis. For example, at time
t=0.75, one stagnation point lies between Z=0.0, Z=0.25 and the other
is situated between Z=0.25 and Z=0.5 which separate the central region
of a retrograde flow from two forward flow regions. Similar
observations had been reported numerically and analytically by
Takabatake and Ayukawa \cite{Takabatake1} and Maiti and Misra
\cite{Maiti1,Maiti2}, Misra and Maiti \cite{Misra2} for a Newtonian
fluid and non-Newtonian fluids respectively for peristaltic
transport. From the standpoint of ciliary pumping, this retrograde
flow at the trough region is considered to be a kind of ineffective
leakage.

\begin{figure}
\centering
\begin{minipage}[b]{0.47\textwidth}
\centering
\subfigure[Velocity profiles at the wave crest]{
\label{fig3.1}
\includegraphics[width=3.6in]{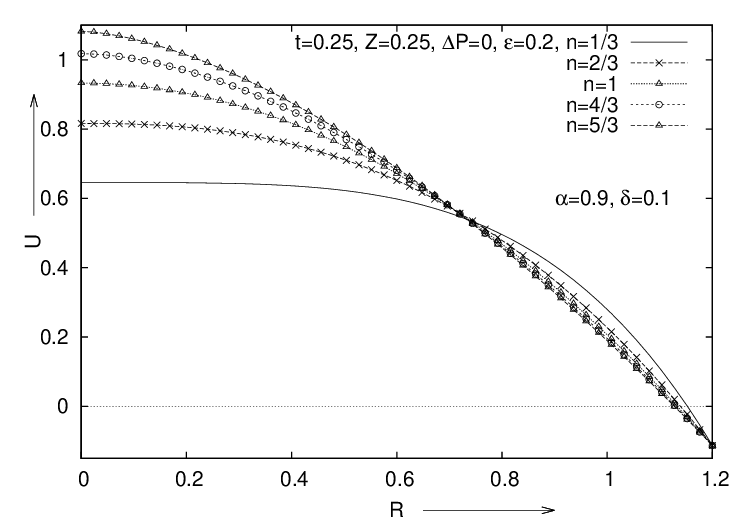}}
\end{minipage}\hspace{0.3in}
\begin{minipage}[b]{0.47\textwidth}
\centering
\subfigure[Velocity profiles at the wave trough]{
\label{fig3.6}
\includegraphics[width=3.6in]{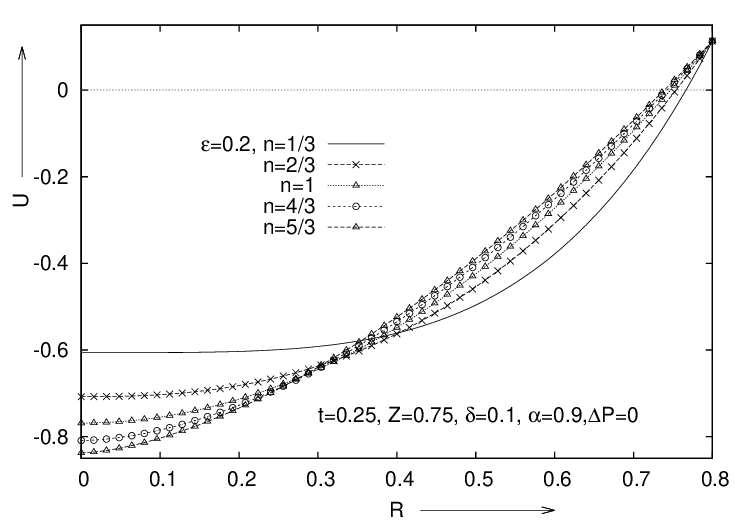}}
\end{minipage}
\caption{The diagrams exhibit the velocity profiles of the
  rheological fluids at the wave crest and the wave trough sections for
  the different values of flow behaviour index $n$.}
\label{manuscript_velo7.3.1-7.3.6}
\end{figure}

It is to be noted that the beat of a single cilium have two different
phases. One of them is the effective stroke of cilia, i.e., when the
movement of cilium is in the general fluid movement direction. Reflux
occurs near the walls in this case and the portion of the tube is
wide. The other phase is the recovery stroke, i.e., when the movement
of cilium is in the opposite direction to the general fluid
movement. There may be forward flow near the walls and the narrow
portion of the tube in this case. Fig. \ref{manuscript_velovect} shows
instantaneous velocity field plot, the corresponding envelop of the
cilia tips is shown in Fig. \ref{manuscript_velo7.1.1-7.1.4}(b). The
flow is along the direction of the cilia motion in the region $0\le Z \le
0.5$, while the reverse trend occurs in the region $0.5\le Z \le 1$.

\begin{figure}
\centering
\begin{minipage}[b]{0.47\textwidth}
\centering
\subfigure[Velocity profiles for a shear thinning fluid at the \newline wave crest]{
\label{fig4.1}
\includegraphics[width=3.6in]{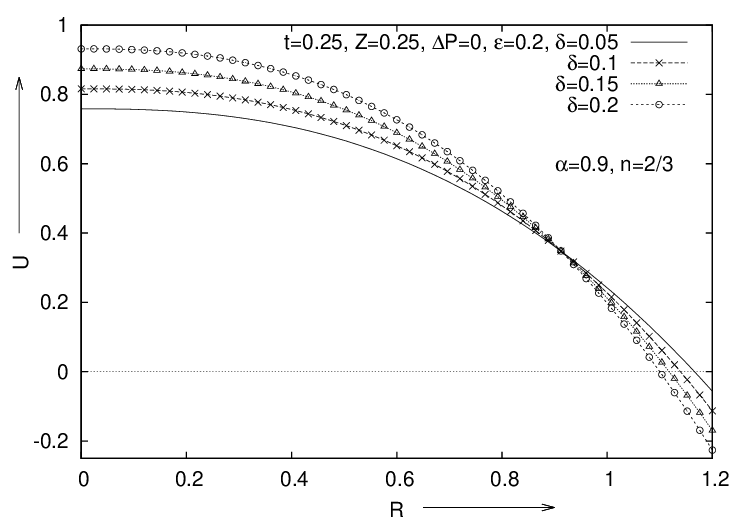}}
\end{minipage}\hspace{0.3in}
\begin{minipage}[b]{0.47\textwidth}
\centering
\subfigure[Velocity profiles for a shear thinning fluid at the \newline wave trough]{
\label{fig4.5}
\includegraphics[width=3.6in]{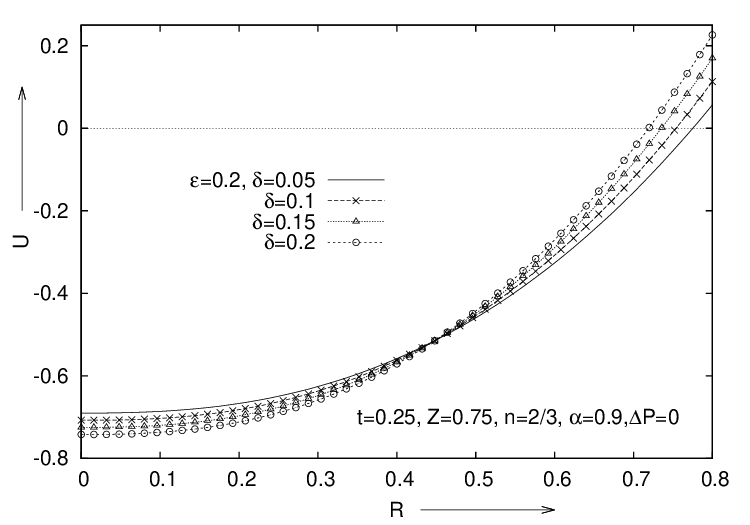}}
\end{minipage}
\\
\begin{minipage}[b]{0.47\textwidth}
\centering
\subfigure[Velocity profiles for a shear thickening fluid at the \newline wave crest]{
\label{fig3.9}
\includegraphics[width=3.6in]{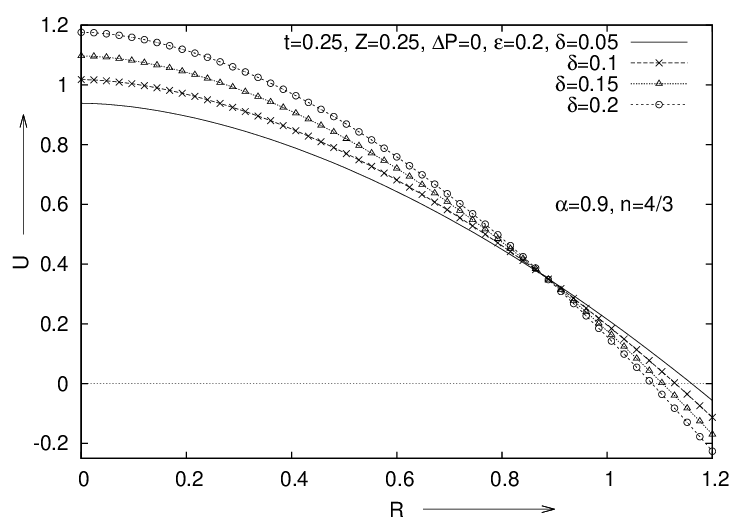}}
\end{minipage}\hspace{0.3in}
\begin{minipage}[b]{0.47\textwidth}
\centering
\subfigure[Velocity profiles for a shear thickening fluid at the \newline wave trough]{
\label{fig4.13}
\includegraphics[width=3.6in]{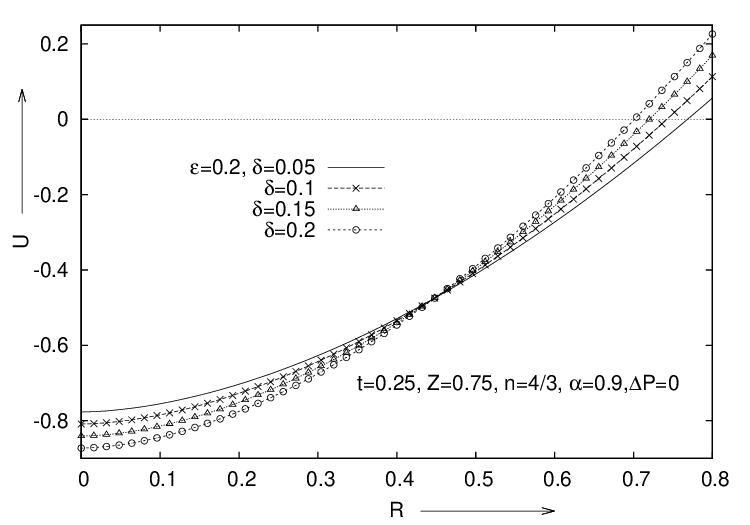}}
\end{minipage}
\caption{The diagrams exhibit the velocity profiles of the
  rheological fluids at the wave crest and the wave trough sections for
  the different values of $\delta$.}
\label{manuscript_velo7.4.1-7.4.13}
\end{figure}

The effects of $\epsilon$, $n$, $\delta$ and $\alpha$ on the
velocities at the crest and the trough section of the tube are shown
in
Figs. (\ref{manuscript_velo7.2.1-7.2.6}-\ref{manuscript_velo7.5.1-7.5.13})
under consideration of free-pumping. It may be observed from the
Fig. (\ref{manuscript_velo7.2.1-7.2.6}) that there is a remarkable
increase in the magnitude of the axial velocity due to an increase in
the value of $\epsilon$ for both sections. This may be interpreted
that the rise of cilium length (at least up to a certain limit) may
generate greater driving
force. Fig. (\ref{manuscript_velo7.3.1-7.3.6}) illustrates that the
magnitude of the axial velocity for both sections enhances
significantly at the central region while the reverse trend occurs at
the boundary region if flow index number $n$ increases. That reverse
trend might have been initiated due to the increase of $n$ which in
related to friction at the boundary region of the tube. As shown in
Figs. (\ref{manuscript_velo7.4.1-7.4.13}-\ref{manuscript_velo7.5.1-7.5.13}),
wave number $\delta$ and eccentricity $\alpha$ of the path of the
cilia raise the magnitude of the axial velocity at the central region
for both sections, while at the boundary region, the trend is reversed
indicating enhancement of friction due to increase in $\delta$ and
$\alpha$.

\begin{figure}
\centering
\begin{minipage}[b]{0.47\textwidth}
\centering
\subfigure[Velocity profiles for a shear thinning fluid at \newline the wave crest]{
\label{fig5.1}
\includegraphics[width=3.6in]{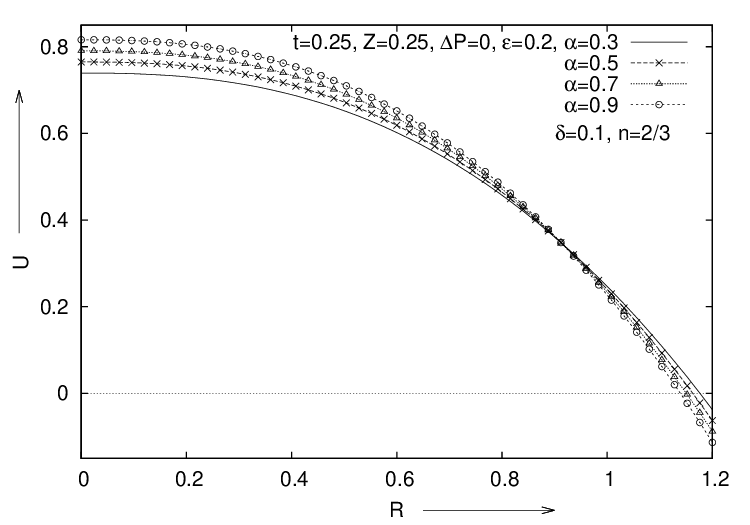}}
\end{minipage}\hspace{0.35in}
\begin{minipage}[b]{0.47\textwidth}
\centering
\subfigure[Velocity profiles for a shear thinning fluid at the \newline wave trough]{
\label{fig3.5}
\includegraphics[width=3.6in]{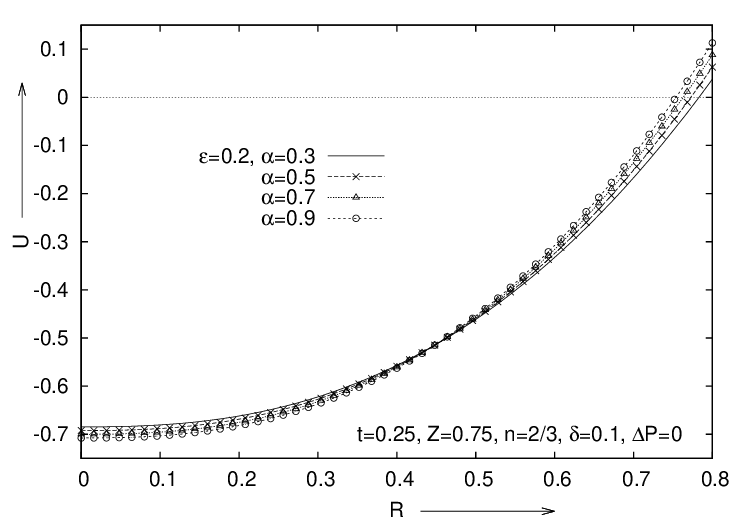}}
\end{minipage}
\\
\begin{minipage}[b]{0.47\textwidth}
\centering
\subfigure[Velocity profiles for a shear thickening fluid at the \newline wave crest]{
\label{fig5.9}
\includegraphics[width=3.6in]{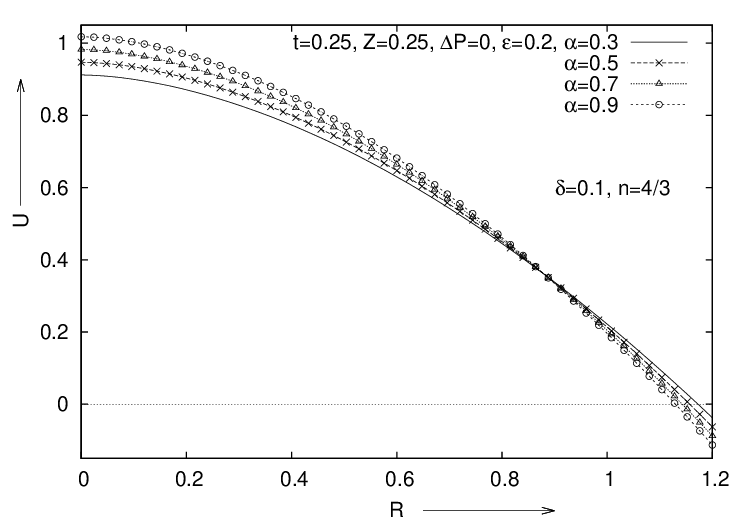}}
\end{minipage}\hspace{0.35in}
\begin{minipage}[b]{0.47\textwidth}
\centering
\subfigure[Velocity profiles for a shear thickening fluid at the \newline wave trough]{
\label{fig5.13}
\includegraphics[width=3.6in]{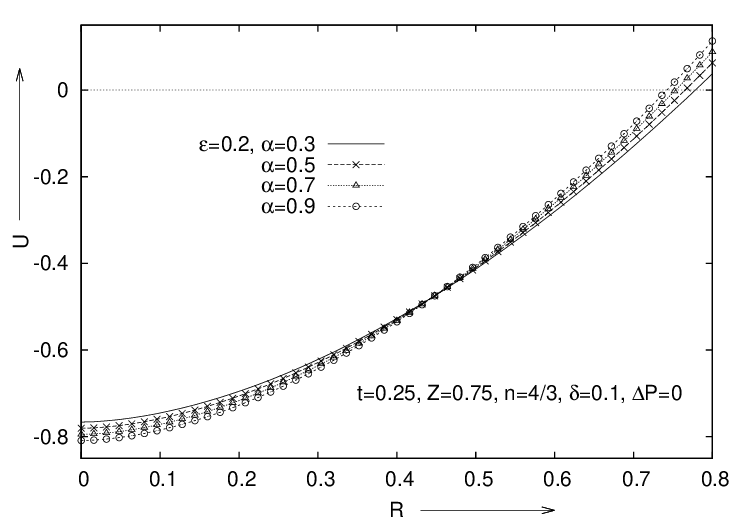}}
\end{minipage}
\caption{The diagrams exhibit the velocity profiles of the rheological fluids at the crest and the
  trough section for the different values of $\alpha$.}
\label{manuscript_velo7.5.1-7.5.13}
\end{figure}

\section{Application to fluid movement in ductuli efferentes} 
The ciliated walls of an axisymmetric tube have been modelled by
metachronal wave of cilia which are equivalent to the wavy walls of
peristaltic transport. There is only a few available data in the
existing literature on the flow rates due to the ciliary activity. The
relationship between the pressure difference and the time-mean-volume
flow rate through an axisymmetric tube is given in
(\ref{manuscript_cilia_pressure_rise_relation}). We have to verify
whether the computations given by
(\ref{manuscript_cilia_pressure_rise_relation}) reflect to the flow
rates measured in efferent ducts of the male reproductive tract. In
humans, the efferent ducts are 10-15 tubules connecting the rete
testis to the epididymis and the lining cells in the tubules are
ciliated. In general researchers believe that the cilia drive the
fluid motion \cite{Greep} through the efferent ducts. Lardner and
Shack \cite{Lardner} estimated the approximate flow rate in human rete
testis to each efferent ducts as $6\times$1 0$^{-3}$ $ml/h$ with
corresponding values $a=50\mu m$, frequency of the cilia beat as
20/sec and $c=(20\rm{~beats/sec})\times 10\mu=200\mu/\rm{sec}$ on the
basis of experimental observations \cite{Setchell,Tuck,Waites} of the
flow rates in the rete testis of rat, ram, and bull. The said
quantities validate the action of long wavelength and small Reynolds
number theory ($Re\ll 1$) in this study. Lardner and Shack
\cite{Lardner} calculated the Non-dimensional and dimensional flow
rate to respectively $2.2\times 10^{-2}$ and
$0.12\times$10$^{-3}$$ml/h$ was evaluated by Lardner and Shack
\cite{Lardner} from their model in the case of free pumping assuming
$\epsilon=0.1$, $\delta=0.1$ and $\alpha=1$. But, if the above said values
are retained in the present study under the consideration $n=1$, this
model gives the non-dimensional flow rate in an axisymmetric tube as
$4.54514\times 10^{-2}$ and consequently, the dimensional flow rate is
$0.257\times$10$^{-3}$ $ml/h$
(cf. Fig. \ref{manuscript_pump7.1.1-7.4.1}a). This value, for an
axisymmetric tube is about twice the value reported by Lardner and
Shack \cite{Lardner} for a two-dimensional channel flow. The reason
for this large difference is that they estimated flow rate initially
for a channel and used it for an axisymmetric case. However,
$\epsilon$ is linearly proportional to the cilia length on the basis
of the assumption and $n$ is linked to the rheological fluid (widely
known as non-Newtonian power-law fluid) in the ductus efferentes of
the male reproductive tract. It has been seen from
Fig. \ref{manuscript_pump7.1.1-7.4.1} that the flow rates increase
significantly with an increase in $\epsilon$. The fact, the flow rates
increase with cilia length, was also reported in
\cite{Blake4}. Moreover, for higher values of $\epsilon$, rheological
fluid index $n$ remarkably affect the fluid transport for adverse
pressure gradient (i.e. at pumping zone). If we consider $n=4/3$,
$\epsilon=0.4$, $\Delta P=0$, $\delta=0.1$ and $\alpha=1$ then the
non-dimensional flow rate and dimensional flow rate are calculated as
$0.581611$ and $3.2889\times$10$^{-3}$ $ml/h$ respectively which is
near to the estimated value $6\times 10^{-3}$ $ml/h$. But authors have
no idea about the actual length of cilia and hence the value of
$\epsilon$, due to non-unavailability of real physiological data of
the concerned variables and parameters used in the analysis. The
length of cilia in ductus efferentes is measured about 5 $\mu m$ in
length in the domestic fowl, 7 $\mu m$ ~m in the guinea-fowl and
quail, and 8 $\mu m$ in the turkey (Meleagris gallopavo).

\section{Summary and Conclusion}
Here, we have presented an analysis for the rheological fluid
transport by means of a sequence of beat of cilia which are in an
array and coordinated in such a way to represent the metachronal
rhythm. The objective of the present study is related to the questions
connecting the understanding of the fluid movement through the
efferent ducts of male reproductive tract of human (\cite{Ilio,Greep})
and the influence of cilia on ovum and sperm movement in fallopian
tubes etc. However, an application of the our results for the flow
rates in efferent ducts of male reproductive tract has been discussed
here. On the basis of the derived analytical expressions, extensive
numerical computations have been carried out. The effects of various
parameters, such as $\epsilon$, $n$, $\delta$, $\alpha$ on pumping
characteristic and velocity distribution are investigated in detail in
the case of an axisymmetric tube flow.

The study reveals that for a particular set of values, say
$\epsilon=0.1$, $\delta=1$, $\alpha=1$, $n=1$, $\Delta P=0$ (as
considered in \cite{Lardner}) flow rate in an axisymmetric tube is
about twice the value reported by Lardner and Shack \cite{Lardner} for
a two-dimensional channel. It further reveals that the flow rate
changes remarkably with $\epsilon$ and $n$. When $\epsilon$ is near
about 0.4, our result for flow rate in human ductus efferentes is
close to the estimated value $6\times 10^{-3}$ $ml/h$ as suggested by
Lardner and Shack \cite{Lardner} based on the experimental
observations for the flow rates in efferent ducts in other animals
e.g. rat, ram, and bull.

The respectable variation between the theoretical and the measured
quantities indicates that the metachronal wave of cilia cannot be
responsible for the total flow rate in efferent ducts and there must
be some other important factors accountable for semen movement.  These
factors may be (cf. Ilio and Hess \cite{Ilio}) (a) contraction of
smooth muscle (b) invariable fluid secretion in seminiferous
epithelium (c) contraction of the myoepithelial layer of seminiferous
tubule as well as tunica albuginea in testis (d) the devoid space
generated due to ejaculation of sperm from the lower tract and by
fluid absorption (e) augmented pressure because of the design of
branching and convergence of ductuli. Another important factor may be
the shape of an envelope of the tips of cilia beat which is different
from the one considered by researchers including us. This motivates us
to study the flow through ductus efferentes in the
future. Consequently, for adequate understanding the mechanism
involved in semen movement in ductus efferentes of male reproductive
tract, further theoretical and experimental investigations are
required. It may be noted that forward flow takes place at
the wider portion of the tube, while backward flow occurs at narrower
portion of the tube as observed earlier
\cite{Misra1,Misra2,Maiti2,Takabatake1}, although forward flow is
dominant for positive time-averaged flow rate.\\

{\bf Acknowledgment:} {\it One of the authors, S. Maiti, is grateful
  to the University Grants Commission (UGC), New Delhi for awarding
  the Dr. D. S. Kothari Post Doctoral Fellowship during this
  investigation.}

\end{document}